\documentclass[
  twocolumn,english,aps,pre,
  superscriptaddress,amsmath,amssymb,floatfix,nofootinbib,longbibliography
]{revtex4-2}

\usepackage{amsthm}
\usepackage{amsfonts}
\usepackage{siunitx}
\usepackage{amsmath}
\usepackage{amssymb}
\usepackage{graphicx}
\usepackage{verbatim}
\usepackage[colorlinks]{hyperref}
\usepackage{tikz}
\usepackage{pgfplots}
\usepackage{adjustbox}
\usepackage{braket}
\usepackage{xcolor}
\usepackage{physics}
\usepackage{amssymb} 
\usepackage{graphicx}
\usepackage{dcolumn}
\usepackage{bm}
\usepackage{mathtools}
\usepackage{hyperref}
\usepackage{mathrsfs}
\usepackage{dashrule}
\usepackage{caption}
\usepackage{subcaption}
\usepackage[version=4,arrows=pgf-filled,
textfontname=sffamily,
mathfontname=mathsf]{mhchem}
\usepackage{comment}

\usepackage[font=small,labelfont=bf,
   justification=justified,
   format=plain]{caption} 
   
\definecolor{linkcolor}{RGB}{0,83,166}
\hypersetup{
  colorlinks = true,
  allcolors = {linkcolor}
}

\begin{document}

\title{Geometry-Induced Domain-Wall Pinning and $\mathbb{Z}_2$ Asymmetry in Nominally One-Dimensional Rydberg Arrays  }

\author{Elijah Pelofske}
\email[]{epelofske@lanl.gov}
\affiliation{Theoretical Division, Quantum \& Condensed Matter Physics, Los Alamos National Laboratory, Los Alamos, NM, USA}
\affiliation{Information Systems \& Modeling, Los Alamos National Laboratory, Los Alamos, NM, USA}
\affiliation{Center for Quantum Computing, Los Alamos National Laboratory, Los Alamos, NM, USA}

\author{Frank Barrows}
\email[]{fbarrows@lanl.gov}
\affiliation{Theoretical Division, Quantum \& Condensed Matter Physics, Los Alamos National Laboratory, Los Alamos, NM, USA}
\affiliation{Center for Quantum Computing, Los Alamos National Laboratory, Los Alamos, NM, USA}
\affiliation{Center for Nonlinear Studies, Los Alamos National Laboratory, Los Alamos, NM, USA}

\author{Cristiano Nisoli}
\email[]{cristiano@lanl.gov}
\affiliation{Theoretical Division, Quantum \& Condensed Matter Physics, Los Alamos National Laboratory, Los Alamos, NM, USA}
\affiliation{Center for Quantum Computing, Los Alamos National Laboratory, Los Alamos, NM, USA}
\affiliation{Center for Nonlinear Studies, Los Alamos National Laboratory, Los Alamos, NM, USA}
\affiliation{Information Science and Technology Institute, Los Alamos National Laboratory, Los Alamos, NM, USA}

\begin{abstract}

We perform experiments on QuEra's neutral-Rydberg-atom Aquila quantum computer, using quasi-adiabatic evolution on 1-dimensional models. We use hardware fine-tuning to balance the measurement statistics close to zero net magnetization, in two different geometric outlines: one closed triangular model with 33 atoms, and one 47-atom square with an engineered atom vacancy. The nature of this processor restricts the geometry of the atoms that can be programmed to a 2D plane. We report effects not predicted by the pure 1D Ising model, such as pinning of domain walls and $\mathbb{Z}_2$ symmetry breaking due to the interplay between van der Waals interactions and the Rydberg blockade. In particular, atoms around the atom vacancy and at the vertices of the 1-dimensional square-outline model strongly prefer the Rydberg state, leading to the effective model having a ferromagnetic bond across the vacancy.

\end{abstract}

\maketitle

\section{Introduction}
\label{section:introduction}

Analogue quantum computers have emerged as quantum architectures that allow for direct experiments on coupled systems of qubits~\cite{zhang2025probing, turner2018weak, serbyn2021quantum, Hartse_2025, bernien2017probing, king2025beyond, King_2021}, within a trend of new analog approaches in the domains of condensed matter, magnetic frustration, and strongly interacting many-body quantum systems~\cite{zhang2025probing, turner2018weak, serbyn2021quantum, Hartse_2025, bernien2017probing, king2025beyond, King_2021, King_2021_spin_ice, Pelofske_2026, barrows2025magneticmemoryhysteresisquantum}. 
In particular, neutral-atom based quantum computers represent a powerful and scalable architecture~\cite{Deutsch_2000, Henriet_2020, Urban_2009, Ga_tan_2009, RevModPhys.82.2313}, in which interacting atoms arranged in space function as interacting qubits to mimic a given quantum Hamiltonian. While the full Hamiltonian describing these atoms is quite complex [Eq.~(\ref{eq:full_hamiltonian})], it is often used to reproduce systems described by simpler models, such as the transverse field Ising model (TFIM). On the other hand, phenomena intrinsic to these machines, such as a fine manipulation of the Rydberg blockade to break $\mathbb{Z}_2$ symmetry, have been exploited to realize ensembles of topological order that would escape the TFIM alone~\cite{Samajdar_2021,Semeghini_2021}. 

To investigate the interplay of these effects and benchmark the feasibility, power, and limits of this approach  we employ 1-dimensional (1D) systems as the simplest model~\cite{balewski2025observationanomalystatisticskibblezurek, rava2025interplayconfinementlocalizationprogrammable, Da__2025, hudomal2025ergodicitybreakingmeetscriticality}. We find that a chain that is topologically one-dimensional does not realize a uniform one-dimensional Hamiltonian when folded into a planar Rydberg array: corners and vacancies reshape the longer-range interactions, pin domain walls, break $\mathbb{Z}_2$ symmetry, and select particular spin configurations. 

\begin{figure}[t!]
    \includegraphics[width=0.75\linewidth]{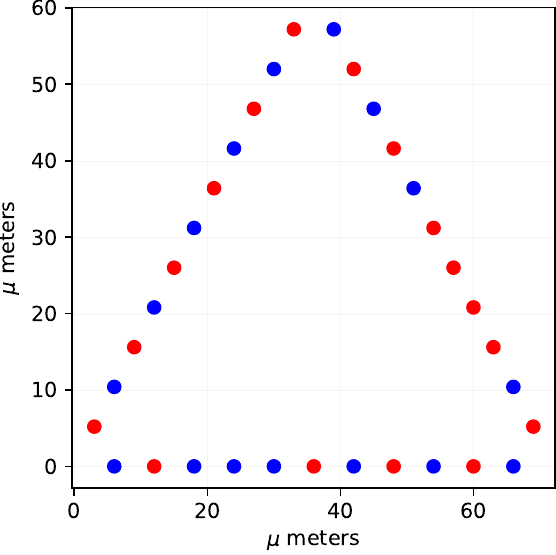}\vspace{3 mm}
    \includegraphics[width=0.75\linewidth]{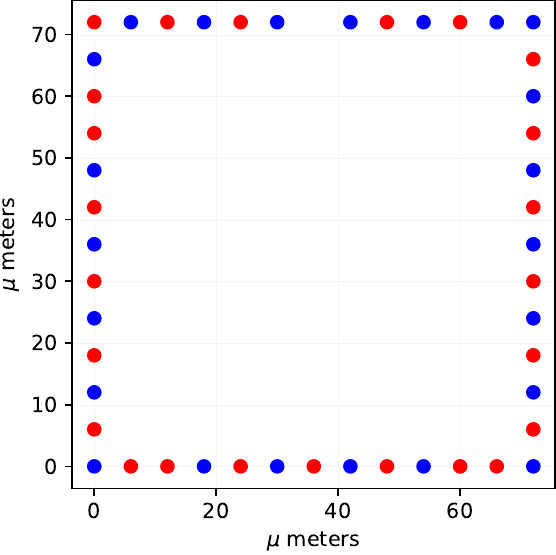}
    \caption{\textbf{The two systems under study.} The position of atoms (dots) as programmed on Aquila for the contiguous triangle outline (top), with $33$ atoms, and for the square outline  (bottom), with $47$ atoms, and an intentional atom vacancy at the top. The axes correspond to real physical space positions, in micrometers. The distance between atoms is $6$ micrometers. Besides the vacancy, the two systems differ at the corners: in the square configuration there are atoms at the corners, but they are absent in the triangular configuration. Both of these atom layouts use the maximum amount of space that the \texttt{Aquila} processor supports, and the atom layouts are restricted to a planar region of space. Colors report the atomic configurations (ground state in red, Rydberg in blue) of a sample measurement chosen to illustrate antiferromagnetic alternation and domain walls. }
    \label{fig:atom_positions}
\end{figure}

We study two 1D arrangements of Rydberg atoms both with the same distance ($6 \mu$m) among Nearest Neighboring (NN) atoms: a closed triangular arrangement and an open square one (Fig.~\ref{fig:atom_positions}), and investigate antiferromagnetic (AFM) domains and their domain walls (DW). 
Because the triangular system has Periodic Boundary Conditions (PBCs), and is composed of an odd number of atoms, only an odd number of DWs is possible, and we find that their statistics is well captured by a Parity Biased (PB) binomial distribution. 
In the open, square outline model one atom is removed, leaving an intentional gap of length corresponding to a double NN distance ($12 \mu$m). The two atoms neighboring the vacancy interact antiferromagnetically with the same strength as the next nearest neighbor (NNN) interaction, which would favor opposite states, the measurement statistics demonstrates their preference for a Rydberg state, and is therefore well captured by a PB binomial distribution that breaks $\mathbb{Z}_2$ symmetry by assuming such preference. We interpret this in terms of lack of the Rydberg blockade at the vacancy leading to an energetic preference of near-vacancy Rydberg atoms. In both arrangements, the spatial distribution of DWs is non-uniform. For the triangular case no atom sits at a corner, and we note that DWs preferentially pin at the corners, where the NNN interactions are stronger and therefore a DW is less costly. In the square case, where an atom sits at each corner, the opposite happens: DWs are pushed away from the corners, whose atoms, again, break $\mathbb{Z}_2$ symmetry by preferring a Rydberg state. This is also reflected in a stronger pinning of the entire qubit texture.

\begin{figure}[t!]
    \centering
    \includegraphics[width=0.89\linewidth]{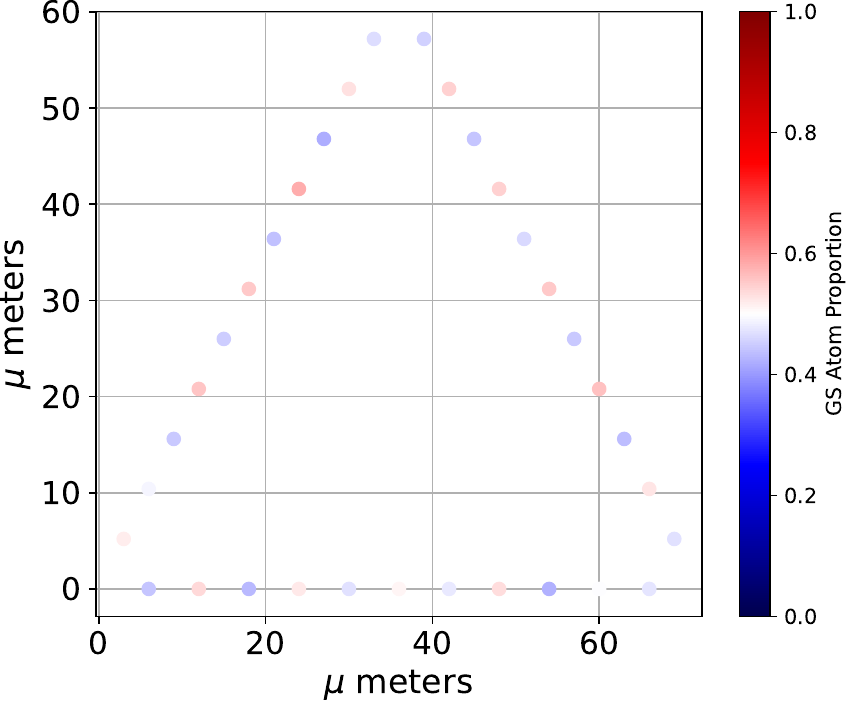}\vspace{3 mm}
     \includegraphics[width=0.89\linewidth]{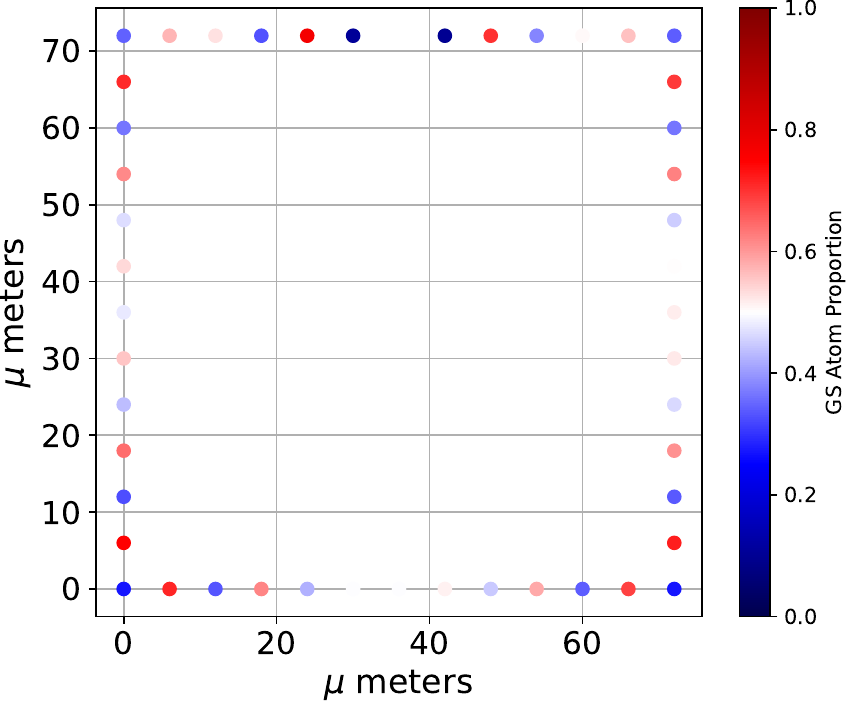}
    \label{fig:samples}
    \caption{{\bf Fraction of Atoms in the Ground state.}  $\mathbb{Z}_2$ symmetry would imply that the fraction is uniformly 0.5. While this is somehow approximated by the triangular case, the square case shows persistence of Rydberg states in the two atoms flanking the vacancy and at the vertices.}
    \label{fig:notz2}
\end{figure}

\section{The Platform and the Experiment}
\label{section:platform}

The \texttt{Aquila} processor~\cite{wurtz2023aquilaqueras256qubitneutralatom} is based on two-state (ground state and Rydberg state) neutral atoms interpreted as qubits coupled by van der Waals interaction. They can be arranged in space by laser beam and interact antiferromagnetically via the Hamiltonian
\begin{align}
    \label{eq:full_hamiltonian}
    \mathcal{H}(t) &= \sum_j \frac{\Omega_j (t)}{2} \left( e^{i \phi} \ket{g_j} \bra{r_j} + e^{- i \phi} \ket{r_j} \bra{g_j}  \right) \nonumber \\
    & - \sum_j \Delta_j (t) \hat{n}_j + \sum_{j < k} V_{jk} \hat{n_j} \hat{n_k}, 
\end{align}
where $\ket{g_j}$ and  $\ket{r_j}$ are the two states (ground state and Rydberg, respectively) of the atom as a qubit ($\ket{g} = \ket{0}$, $\ket{r} = \ket{1}$ and $n=\ket{1}\bra{1}$), $\Omega$ is the Rabi frequency, which drives state transitions in the system, $\phi$ is the phase, and $\Delta$ is the laser detuning. Each of these three is a programmable time-dependent waveform. The Rydberg interaction is then given by  $V_{jk} = {C_6}/{|\vec{r}_j - \vec{r}_k|^6}$
where $C_6 = 5{,}420{,}503~\mu$m$^6$ rad / $\mu$s~\cite{wurtz2023aquilaqueras256qubitneutralatom}.
Key to the Aquila hardware is the Rydberg blockade, for which two atoms within each other's blockade radius are entangled, defining the two-body terms of the quantum Hamiltonian. Different types of quantum Hamiltonians can then be programmed by choosing the atom positions and the three time-dependent waveforms. 

We treat atoms in their ground state as a spin down, and atoms in the excited Rydberg state as a spin up, or $\ket{g} =\ket{\downarrow}, \ket{r} =\ket{\uparrow}$, thus realizing qubits. 
We implement a ``quasi-adiabatic'' ramp  described in refs.~\cite{Ebadi_2022, lukin2024quantumquenchdynamicsshortcut, bernien2017probing, Semeghini_2021}, and ramp $\Delta$ from negative to positive values (this carries out the adiabatic ramp), while holding $\Omega$ at a large fixed value (this facilitates state transitions). The goal of this protocol is to implement a noisy-hardware version of adiabatic quantum evolution for sampling the ground-state configuration of a classical Hamiltonian, such as an Ising model or a combinatorial optimization problem~\cite{farhi2000quantumcomputationadiabaticevolution, PhysRevE.58.5355, kato1950adiabatic, born1928beweis}. We use the longest simulation time programmable on Aquila, i.e. $4$ microseconds, because adiabatic evolution benefits from longer evolution time.

We study the square and triangular 1-dimensional systems of atoms described above to probe the spatial density of its domain walls. From $V_{jk}$ in Eq.~(\ref{eq:full_hamiltonian}), the ground state is an antiferromagnetic configuration of alternating up/down spins whose excitations are DWs of parallel spins that locally break the alternation. To this end, we implement the model via two different geometric arrangements of the neutral atoms, triangular and square, whose exact atom positions are reported in Fig.~\ref{fig:atom_positions} and which we had described above. In principle, a circular arrangement would be preferable because it would exclude non-uniform further neighbor interactions. On \texttt{Aquila} atoms are constrained to a square optical grid of definition $4\time4~\!\mu\text{m}^2$, with an additional constraint that all atoms must be aligned on the same grid in the vertical axis with the minimum spacing of $4~\mu$m. The minimum spacing allowed by the machine is therefore $4~\mu$m, but we chose $6~\mu$m to prevent the Rydberg blockade from dominating~\cite{wurtz2023aquilaqueras256qubitneutralatom}. 
\begin{figure}[t!]
    \centering
    \includegraphics[width=0.99\linewidth]{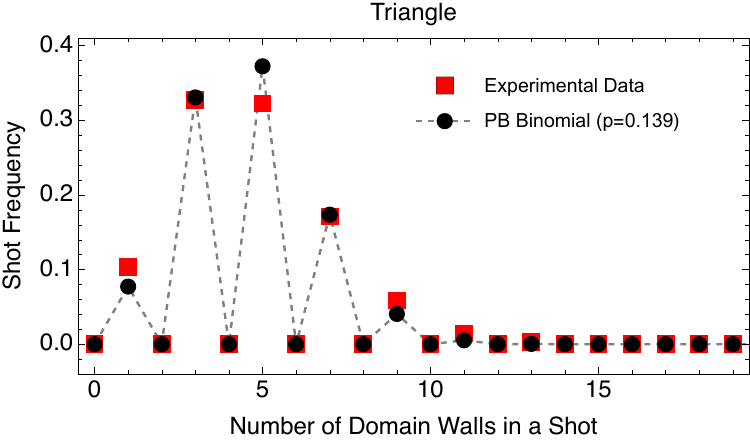}\vspace{2 mm}
    \includegraphics[width=0.99\linewidth]{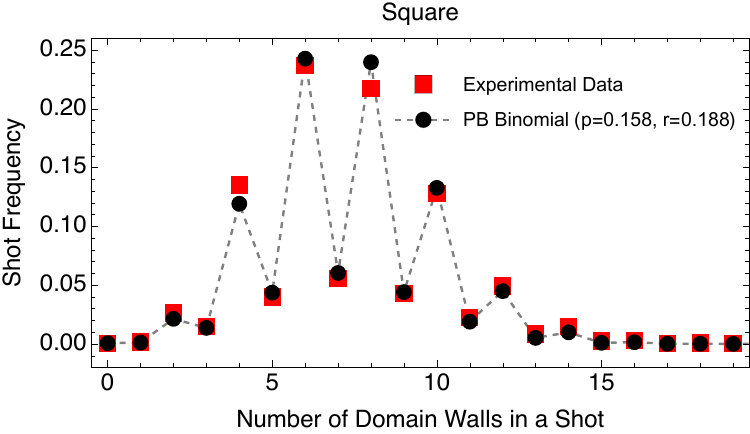}
    \caption{{\bf Domain-wall statistics.} Frequencies of shots vs. number of domain walls in a shot for the triangular (top) and square (bottom) arrangements. In red squares we report the data, and in black dots the corresponding Gibbs distribution.}
    \label{fig:domain_wall_frequency_density}
\end{figure}

Importantly, for both systems, we independently tune the analog schedule parameters such that the proportion of Rydberg atoms (or ``up spins") and ground-state measured atoms (or ``down spins") is approximately in an even ratio to each other, which is what we would expect from antiferromagnetic ordering. For the triangle system, we use a $\Delta$ ramp from $-125$ to $125$ $\mathrm{rad}/\mu\mathrm{s}$ over the course of the $4$ microseconds. We use a Rabi oscillation strength of $\Omega=15.8$ $\mathrm{rad}/\mu\mathrm{s}$, which is the maximum value programmable on the hardware. $\Omega$ is programmed by ramping up from $\Omega=0$ at $t=0$ to the maximum value $\Omega=15.8$ at $0.1$ microseconds, and then ramping down $\Omega$ to zero in the interval from $3.9$ to $4$ microseconds. The ramping of $\Omega$ is a machine requirement because the Rabi oscillations must be at zero both at the beginning and at the end of the simulation. For the square system, we use a $\Delta$ ramp from $-114$ to $114$ $\mathrm{rad}/\mu\mathrm{s}$ over the course of the $4$ microseconds and a $\Omega = 13.0$ $\mathrm{rad}/\mu\mathrm{s}$. For all \texttt{Aquila} experiments, we program the hardware parameters using \texttt{Bloqade.jl}~\cite{bloqade2023quera}, a Julia-based~\cite{Julia-2017} library to define and simulate analog Rydberg atom experiments.

\begin{figure*}[ht!]
    \centering
    \includegraphics[width=0.24\linewidth]{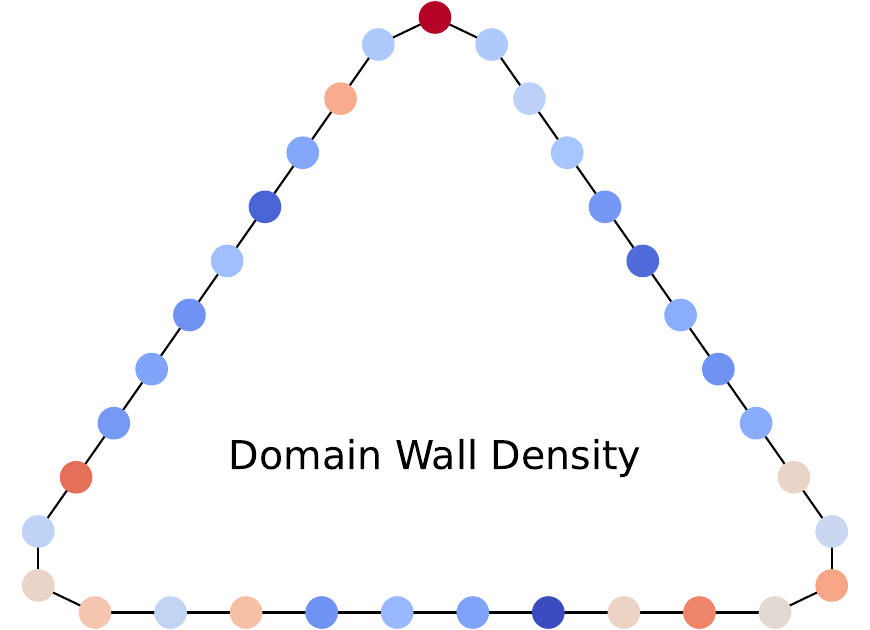}
    \includegraphics[width=0.24\linewidth]{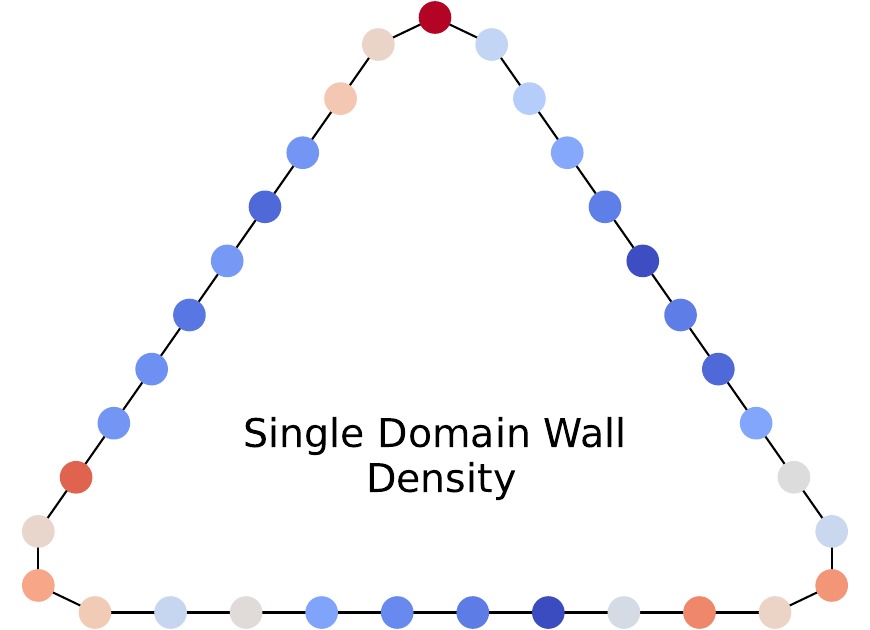}
    \includegraphics[width=0.24\linewidth]{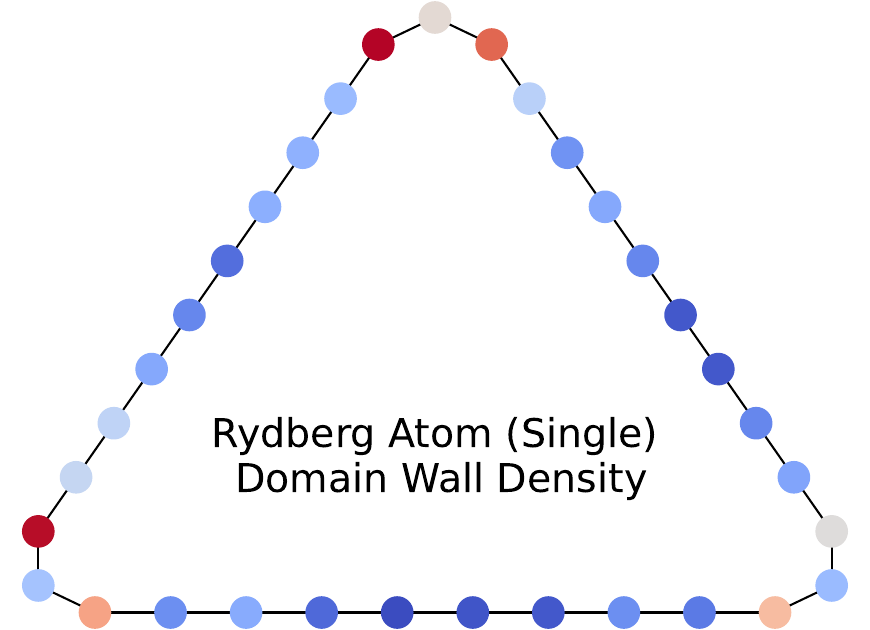}
    \includegraphics[width=0.24\linewidth]{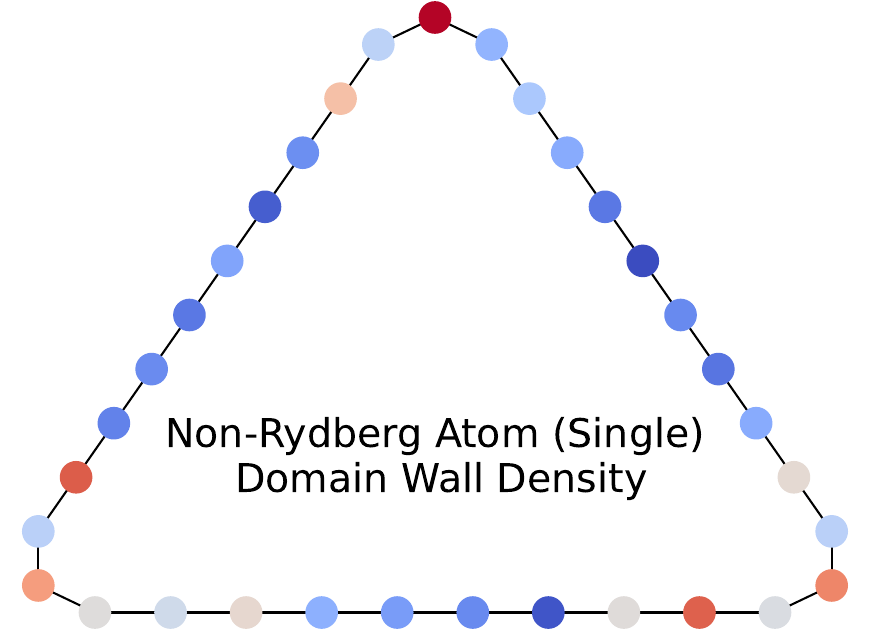}
    \includegraphics[width=0.24\linewidth]{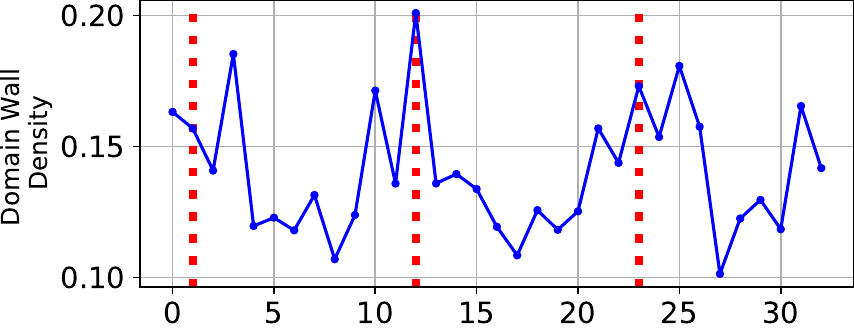}
    \includegraphics[width=0.24\linewidth]{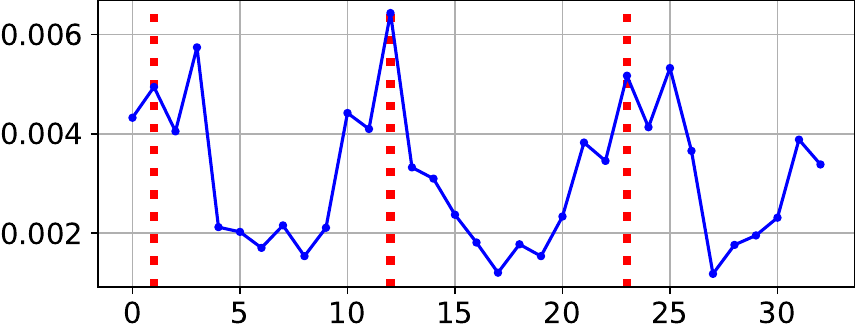}
    \includegraphics[width=0.24\linewidth]{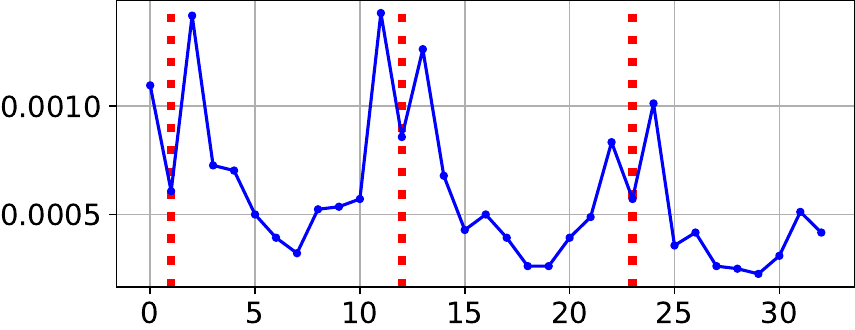}
    \includegraphics[width=0.24\linewidth]{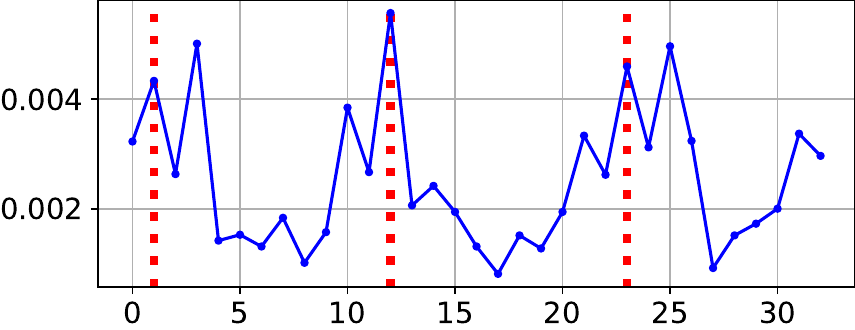}
    \caption{\textbf{Antiferromagnetic domain wall density on the \texttt{Aquila} device on a PBC ring with 33 atoms.} 
    The top row shows graph representations of the domain wall densities where red denotes higher density, blue denotes lower density, and each node position encodes the median position of the bond between two atoms. Top-left: density of all domain walls where each node represents a bond in the 1D model, middle-left: post-selected samples with only a single domain wall, middle-right: count of Rydberg atom (single) domain wall densities, right: count of ground state atom (single) domain wall densities. Due to geometric frustration and the PBCs there is always at least one pinned domain wall somewhere in the chain. 
    The bottom row reports domain wall proportion as a function of edge index where the beginning index is chosen arbitrarily and the dashed vertical red lines denote the bond index at each of the three corners. The domain wall peaks correspond approximately to the corners of the triangle layout. 
    }
    \label{fig:1D_triangle_outline_AFM_domain_wall_density}
\end{figure*}

\section{Results}
\label{section:results}

We execute a total of $100{,}000$ shots on the \texttt{Aquila} processor for each of the two systems. The primary source of error on the \texttt{Aquila} processor is measurement error, see Ref.~\cite{wurtz2023aquilaqueras256qubitneutralatom} for calibrated error rates. Another source of error comes from a non-zero probability (approximately $0.007$, but is geometry dependent) that atom positions that were programmed to be ``filled'' fail to be filled on the hardware at the time of the experiment~\cite{wurtz2023aquilaqueras256qubitneutralatom}. Of the $100{,}000$ samples for the triangle system, $83{,}950$ were retained after post-selection that eliminated shots in which atoms were not filling correctly; the rest of these samples were discarded due to the atom filling error which corresponds to failure of the machine to realize the desired geometry. Of these, $8{,}659$ of the samples had a single domain wall, which corresponds to the ground state of the triangle as an AFM closed, odd-spin system. For the square system, of the $100{,}000$ shots only $66{,}374$ samples were kept, with the rest being discarded due to having a missing atom in the beginning configuration of the array.

The colors in Fig.~\ref{fig:atom_positions} demonstrate two examples of the outcomes of shots for both the triangular and square geometry. In Fig.~\ref{fig:notz2} we plot a heatmap corresponding to the average spin for the triangular and square outline. If the system were a pure Ising model, $\mathbb{Z}_2$ symmetry would imply that all the average spins are zero. In the triangular case, we indeed see close to zero net magnetization, resulting in a fainter heatmap. Interestingly, however, we see a clear alternation of the average magnetization on each site, which is due to stronger NNN interactions pinning DWs at the three corners. In the square case $\mathbb{Z}_2$ symmetry is broken substantially more than in the triangular case, where corner effects at the four corners, along with the atom vacancy, cause a clear antiferromagnetic ordering to emerge on average. The square case demonstrates a frustration of the Rydberg blockade at the four corners, which form triangles within which it is only energetically favorable for one atom to be in the Rydberg state, which in all four cases on average is the single corner atom. Importantly, we note that in the square outline model, the two atoms flanking the vacancy are preferentially in the Rydberg state. While these atoms weakly interact antiferromagnetically, the Rydberg effect brought by the vacancy keeps them preferentially in a Rydberg state.

In Fig.~\ref{fig:domain_wall_frequency_density} we report the domain wall statistics. The triangular outline--a closed antiferromagnetic system with an odd number of qubits $N_{\text{qb}}=33$--cannot host an even number of domain walls. The ensemble is well described by a Parity Biased (PB) Binomial distribution $p_n\propto \binom{N_{\text{qb}}}{n}p_{\triangle}^n (1-p_{\triangle})^{N_{\text{qb}}-n}$ for $n$ odd, and $p_n=0$ for $n$ even. $p_n$ is the probability of obtaining $n$ domain walls and $p_{\triangle}$ is the probability of a domain wall. By minimizing the Kullback–Leibler divergence between the PB Binomial distribution and the data, we obtain a good fit (top of Fig.~\ref{fig:domain_wall_frequency_density}) for $p_{\triangle}\simeq 0.14$.

The square case is more interesting. Being an open-boundary system it supports both even and odd numbers of domain walls. However, the bottom plot in Fig.~\ref{fig:domain_wall_frequency_density} shows that odd numbers of domain walls are suppressed.
Because the number of atoms $N_{\text{qb}}=47$ is odd, an odd number of domain walls would result in an antiferromagnetic arrangement of the two atoms around the vacancy, and such a configuration should be slightly favored by the NNN antiferromagnetic interaction, rather than suppressed. Instead, Figs.~\ref{fig:domain_wall_frequency_density} and \ref{fig:notz2} tell us that the system not only greatly favors a ferromagnetic orientation among those two spins, but in fact the Rydberg orientation. Indeed, the data is well fitted by PB Binomial distribution that ascribes biased probabilities to odd and even domain walls--or $p_n\propto \binom{N_{\text{qb}}}{n}p_{\square}^n (1-p_{\square})^{N_{\text{qb}}-n} r$ for $n$, $p_n\propto \binom{N_{\text{qb}}}{n}p_{\square}^n (1-p_{\square})^{N_{\text{qb}}-n}(1-r)$ for $n$ even. By minimizing the Kullback–Leibler divergence between the data and the PB Binomial distribution we find $p_{\square}\simeq 0.16$ (not too far from $p_{\triangle}\simeq 0.14$) and $r\simeq 0.19$ for the bias that suppresses odd domain walls.

We turn now our attention to Fig.~\ref{fig:1D_triangle_outline_AFM_domain_wall_density} which displays the measured DW spatial densities for the triangular system. Interestingly, they are not uniform, as one would expect in a 1D Ising model. Once again, the case of the closed triangle is simpler: the domain wall peaks are pinned approximately to the corners of the triangle outline. As discussed in Supplementary Information~\ref{section:appendix_NNN}, the non-uniform next-nearest-neighbor interaction at the corners ($V_\text{NNN}^{(\text{corner})} \approx 2.4\, V_\text{NNN}^{(\text{edge})}$)--perhaps also compounded by spurious couplings~\cite{tüysüz2025learningresponsefunctionsanalog}--lifts the translational degeneracy creating an energy landscape that favors wall localization at the three corners. Because of that, the profiles in Fig.~\ref{fig:1D_triangle_outline_AFM_domain_wall_density} demonstrate a robust pinning of DWs, whereas Fig.~\ref{fig:notz2} shows only a faint pinning of qubits. 

\begin{figure}[ht!]
    \centering
    \includegraphics[width=0.494\linewidth]{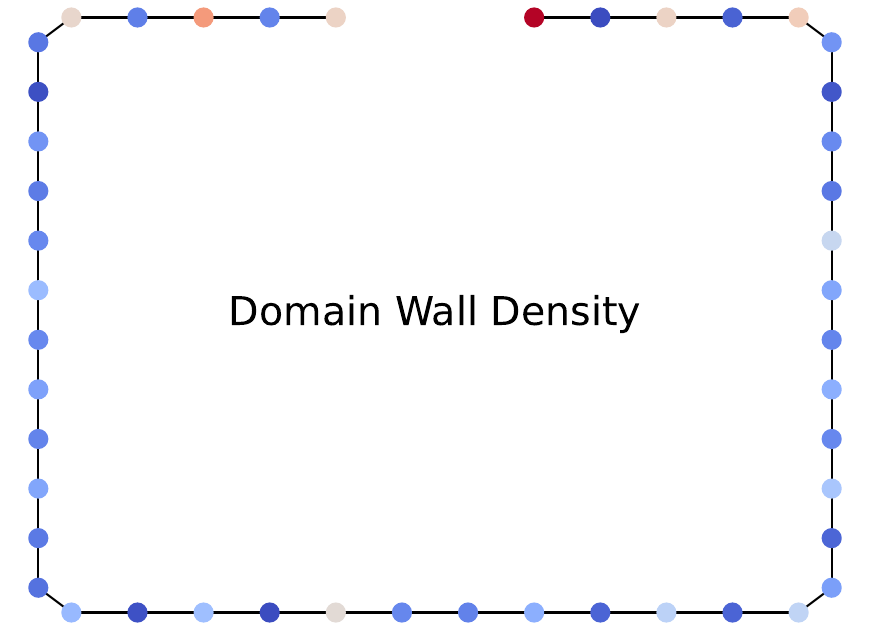}
    \includegraphics[width=0.494\linewidth]{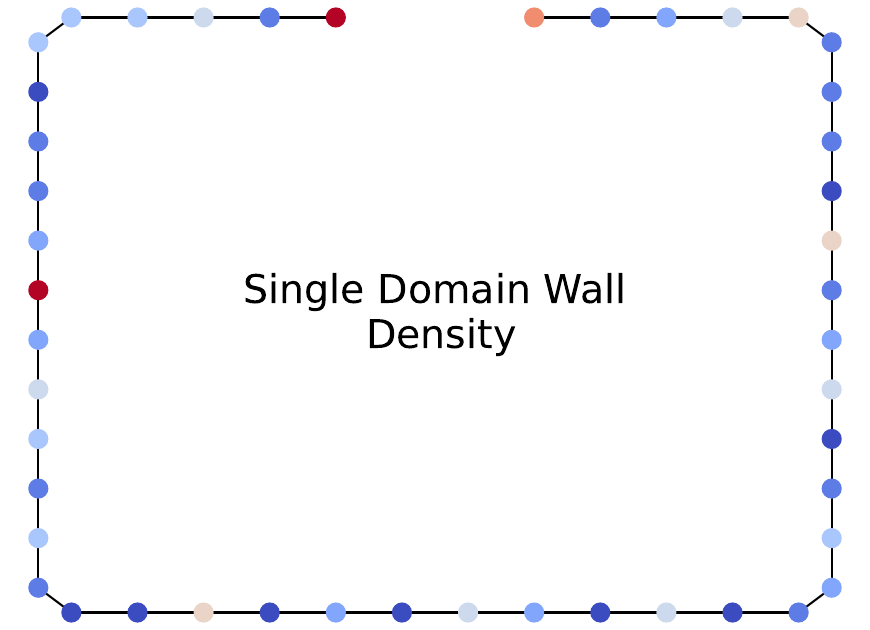}\\
    \includegraphics[width=0.494\linewidth]{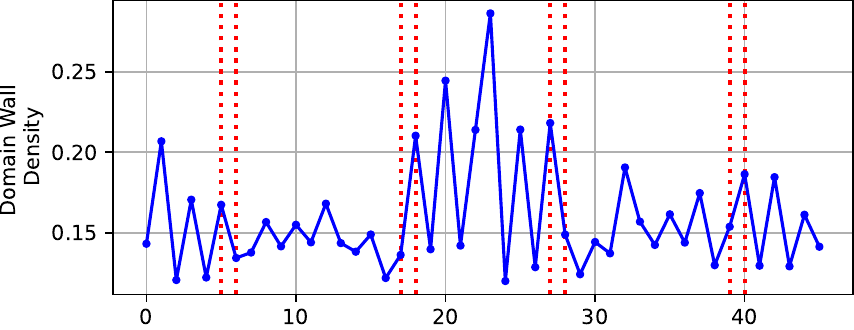}
    \includegraphics[width=0.494\linewidth]{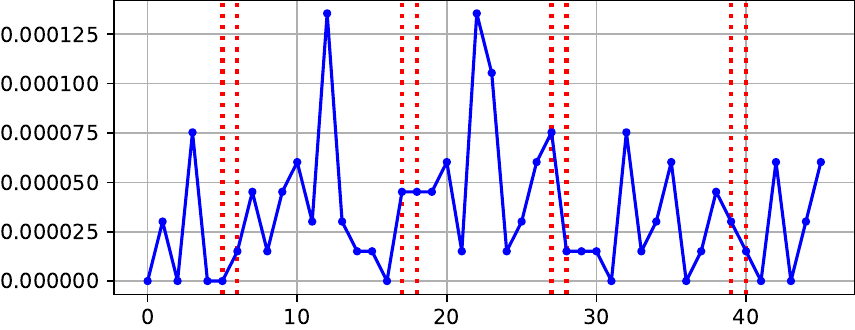}
    \caption{ \textbf{Domain wall densities on the 47-atom 1D square outline simulations with an intentional atom vacancy. } 
    The vertical dashed lines denote the (two) bonds that form the angles in the square. The two center bonds in this plot correspond to the two bonds on either side of the engineered atom vacancy; the site of the missing atom is not counted as a bond in the data. 
    }
    \label{fig:1D_sqaure_outline_AFM_domain_wall_density}
\end{figure}

Next, Fig.~\ref{fig:1D_sqaure_outline_AFM_domain_wall_density} plots the domain wall density of the square model, where we note that DWs reside away from the boundaries. The density contains substantially more domain wall oscillations compared to the triangular model domain wall densities of Fig.~\ref{fig:1D_triangle_outline_AFM_domain_wall_density}. The oscillations come from a combination of pinning of Rydberg atoms at the corners of the model (see also Fig.~\ref{fig:notz2}), and adjacent to the vacancy, combined with the odd-side-lengths on the model, thus leading to frustration on each of the sides of the square. In order to reveal antiferromagnetic domains, Supplementary Information~\ref{section:appendix_AFM_order_param_samples} plots single-measurement examples of antiferromagnetic domains revealed with a staggered antiferromagnetic order parameter applied to the square model.

As a consistency check on our initial calibration of adjusting the $\Delta$ detuning  strength and atom spacing, we evaluated how close to parity the aggregate \texttt{Aquila} measurement statistics are. With the aim of implementing effective antiferromagnetic Ising models, sampling close to an even distribution of qubits in the Rydberg state and the ground state is ideal. To this end, our initial calibration succeeded: for the triangle 1D model a total of $2{,}770{,}350$ qubit measurements were collected (after post-selection) of which $50.867\%$ were in the Rydberg state. For the open boundary system model, a total of $3{,}119{,}578$ qubit measurements were collected of which $50.790\%$ were in the Rydberg state. This shows that the results we present are as close as possible to an Ising-like simulation, in terms of net magnetization. 

\section{Discussion}
\label{section:discussion}

The Rydberg atom encoding is widely used for Maximum Independent Set optimization~\cite{Ebadi_2022, pichler2018quantumoptimizationmaximumindependent} and for probing Kagome lattice ground states~\cite{Samajdar_2021}, but it does not possess the $\mathbb{Z}_2$ symmetry of the antiferromagnetic Ising model. The detuning term $-\Delta\hat{n}_j$ assigns different energies to the $\ket{r}$ and $\ket{g}$ states, and the blockade interaction $V_{jk}\hat{n}_j\hat{n}_k$ penalizes adjacent $\ket{r}$ pairs while placing no penalty on adjacent $\ket{g}$ pairs. In the PXP description of these dynamics~\cite{turner2018weak, serbyn2021quantum}, this asymmetry is absolute: $\uparrow\uparrow$ domain walls ($\ket{r_j}\ket{r_{j+1}}$) are excluded from the constrained Hilbert space entirely, and only $\downarrow\downarrow$ walls ($\ket{g_j}\ket{g_{j+1}}$) are permitted.

In the limit where $V_\text{NN} \gg \Omega, \Delta$, the dynamics are confined to a constrained Hilbert space in which no two neighboring atoms simultaneously occupy $\ket{r}$. The effective Hamiltonian in this subspace is
\begin{equation}
    H_\text{PXP}
    = \frac{\Omega}{2}\sum_j P_{j-1}\,\sigma^x_j\,P_{j+1}
      - \Delta\sum_j \hat{n}_j,
    \label{eq:pxp}
\end{equation}
where $P_j = 1 - \hat{n}_j$ projects onto $\ket{g}$ at site $j$. The projectors enforce the blockade constraint: spin flips at site $j$ are permitted only when both neighbors are in $\ket{g}$. This contrasts with a conventional transverse-field Ising model, which possesses full $\mathbb{Z}_2$ symmetry between $\uparrow\uparrow$ and $\downarrow\downarrow$ walls.

The domain wall statistics reveal this asymmetry between the two types of antiferromagnetic domain walls. Among the post-selected single-domain-wall samples, $\downarrow\downarrow$ walls (two adjacent atoms in $\ket{g}$) outnumber $\uparrow\uparrow$ walls (two adjacent atoms in $\ket{r}$) by approximately $3{:}1$ (see the two rightmost columns of Fig.~\ref{fig:1D_triangle_outline_AFM_domain_wall_density}).

For the experimental parameters, the blockade ratio is $V_\text{NN}/\Omega \approx 116/15.8 \approx 7.4$ (triangle) and $\approx 116/13.0 \approx 8.9$ (square). These values are large but finite, so the PXP description is a good approximation but not exact. We can also compare the energetics of the two wall types directly. For the PBC triangle with $N=33$ and a single domain wall, the $\downarrow\downarrow$ configuration contains $(N-1)/2 = 16$ Rydberg excitations with no adjacent Rydberg pairs, while the $\uparrow\uparrow$ configuration contains $(N+1)/2 = 17$ excitations with one adjacent pair at the wall. Neglecting the (small) next-nearest-neighbor contributions, the energy difference in the classical limit ($\Omega \to 0$) is
\begin{equation}
    E_{\uparrow\uparrow} - E_{\downarrow\downarrow}
    \approx -\Delta + V_\text{NN}.
    \label{eq:dw_split}
\end{equation}
At the final detuning $\Delta_\text{final} = 125\;\text{rad}/\mu\text{s}$ and $V_\text{NN} \approx 116\;\text{rad}/\mu\text{s}$, this gives $E_{\uparrow\uparrow} - E_{\downarrow\downarrow} \approx -9\;\text{rad}/\mu\text{s}$: the two wall types are nearly degenerate, with $\uparrow\uparrow$ marginally lower in energy.

Despite the near-degeneracy (or possible energetic preference for $\uparrow\uparrow$), the experiment strongly favors $\downarrow\downarrow$. This is a result of the constrained dynamics during the quasi-adiabatic ramp. The evolution begins in the trivial ground state (all $\ket{g}$ at large negative $\Delta$) and proceeds through a regime where $\Omega$ is large and the PXP constraint is well enforced.

\section{Conclusion}
\label{section:conclusion}

We have presented experiments run on the QuEra \texttt{Aquila} neutral-atom quantum computer using quasi-adiabatic analog ramp protocols to probe antiferromagnetic domain walls in two nominally one-dimensional Rydberg arrays embedded on a 2D plane: a closed 33-atom triangular chain and a 47-atom square outline interrupted by a single vacancy. At the level of domain-wall-number statistics, the data are captured by parity biased binomial distributions. The qubit measurement spatial distributions, both of site-magnetization and domain wall densities, are strongly nonuniform. In the triangle, the enhanced next-nearest-neighbor interactions at the three corners lift the translational degeneracy of the nearest-neighbor Ising-like manifold and pin the domain walls. In the square, the corners and vacancy produce a much stronger site-resolved texture, with higher Rydberg atom measurement probability at the vertices and at the two atoms adjacent to the vacancy.

These observations show that a geometrically one-dimensional Rydberg array whose net magnetization has been carefully tuned using the analog hardware controls to be approximately zero, does not produce uniform one-dimensional Ising Hamiltonian statistics at the level of site-resolved magnetization, or domain wall densities. Our results characterize how non-uniform atom placement in the 2D plane and atom vacancies can select spin configurations, bias domain-wall parity, violate $\mathbb{Z}_2$ symmetry and localize defects even when the nearest-neighbor spacing is uniform.

The quantum evolution with the Rydberg blockade also leaves a signature in the nature of the defects. Among the single domain wall configurations of the triangular PBC ring, $\lvert gg \rangle$ DWs occur approximately three times more frequently than $\lvert rr \rangle$ DWs, although a nearest-neighbor energy penalty estimate makes the two wall types nearly degenerate at the end of the ramp. This imbalance is consistent with the energy penalty of $\lvert rr \rangle$ DWs in the Rydberg blockade.

Finally, we find that the sampling rate of ground state configurations of the effective 1D antiferromagnet model is relatively low. The frequent production of multiple walls is consistent with a finite-time ramp ($4$ microseconds), at the given hardware energy scale, leading to non-adiabatic defect creation. DW creation in these experiments could also be increased by hardware imperfections such as analog control errors or biased qubit decoherence.

\section*{Acknowledgments}
\label{sec:acknowledgments}
We thank Carleton Coffrin for valuable discussions. 
This work was supported by the U.S. Department of Energy through the Los Alamos National Laboratory. Los Alamos National Laboratory is operated by Triad National Security, LLC, for the National Nuclear Security Administration of the U.S. Department of Energy (Contract No. 89233218CNA000001). The research presented in this article was supported by the Laboratory Directed Research and Development program of Los Alamos National Laboratory under project number 20240032DR. This research used resources provided by the Los Alamos National Laboratory Institutional Computing Program. The authors would also like to thank the New Mexico Consortium, under subcontract C2778, the Quantum Cloud Access Project (QCAP), for providing quantum computing resources. LA-UR-26-22772.

\section*{Author Contributions}
EP, FB and CN conceived of the project, EP performed the experiments on Aquila, FB and CN supervised the project, all authors contributed to the data analysis, and all authors contributed to the final manuscript. 

\appendix

\section{Next Nearest Neighbor Interactions}
\label{section:appendix_NNN}

Next-nearest-neighbor and longer-range interactions cannot be removed from the dynamics of \texttt{Aquila}. We quantify the scale of these residual interactions for the geometries used here. With a uniform nearest-neighbor spacing $a = 6\;\mu$m, the leading interaction is
\begin{equation}
    V_\text{NN} = \frac{C_6}{a^6}
    = \frac{5{,}420{,}503}{6^6}
    \approx 116.2\;\text{rad}/\mu\text{s}.
    \label{eq:vnn}
\end{equation}
Along a straight edge the next-nearest-neighbor separation is $2a$, giving 
\begin{equation}
V_\text{NNN}^{(\text{edge})} = V_\text{NN}/64 \approx 1.82\;\text{rad}/\mu\text{s}.
\end{equation}
At the clipped corners of the triangle outline, the geometry is different. The corner vertex atom has been removed, and the two flanking atoms sit on edges that meet at $60^\circ$. Their  separation remains
$a=6\;\mu$m, as throughout the rest of the system. The next-nearest-neighbor distance across a corner, however, is $\sqrt{3}a
    \approx 10.39\;\mu\text{m}$
which is shorter than the straight-edge value of $12\;\mu$m. The corresponding interaction is \begin{equation}
    V_\text{NNN}^{(\text{corner})} = V_\text{NN}/27 \approx 4.30\;\text{rad}/\mu\text{s}, 
\end{equation}
a factor of $64/27 \approx 2.4$ stronger than  $V_\text{NNN}^{(\text{edge})}$. 

For a 1D Ising model  with only nearest-neighbor interactions, all positions for a single antiferromagnetic domain would be equivalent. Our triangular system breaks this symmetry for the NNN coupling. A stronger antiferromagnetic NNN interaction slightly favors domain walls at the vertices, with an energy splitting set by the scale
$V_\text{NNN}^{(\text{corner})} - V_\text{NNN}^{(\text{edge})}
\approx 2.5\;\text{rad}/\mu\text{s}$.
Similarly, the $90^\circ$ corners of the square outline introduce a next-nearest-neighbor distance of $a\sqrt{2} \approx 8.49\;\mu$m, giving $V_\text{NNN}^{(\text{sq.\,corner})} = V_\text{NN}/8 \approx 14.5\;\text{rad}/\mu\text{s}$, which is an even  larger perturbation to the uniformity of the system than in the triangular case.

\section{Staggered Antiferromagnetic Order Representative Samples}
\label{section:appendix_AFM_order_param_samples}

\begin{figure}[ht!]
    \centering
    \includegraphics[width=0.32\linewidth]{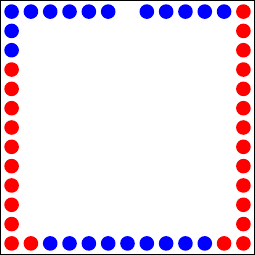}
    \includegraphics[width=0.32\linewidth]{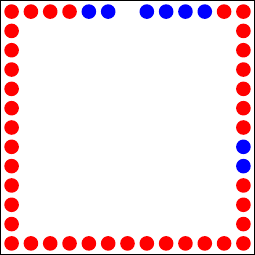}
    \includegraphics[width=0.32\linewidth]{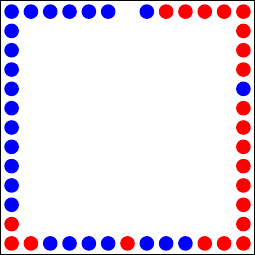}
    \caption{ \textbf{Representative qubit measurements, with a staggered Antiferromagnetic order parameter applied.} Atom measurements of the 1D square case, with a staggered antiferromagnetic order parameter applied so as to highlight antiferromagnetic domains. }  
    \label{fig:samples_AFM_order_param}
\end{figure}

Fig.~\ref{fig:samples_AFM_order_param} reports example measurements, with a staggered spin sign reversal function applied $\tau_{i} =(-1)^{i}\,\sigma_{i}^z$, on the square 1D chain. We do not show the same representation for the triangle 1D case because in that case the degeneracy and PBCs make such an order parameter ill-defined.

\bibliographystyle{apsrev4-2-titles}
\bibliography{references}
\end{document}